\documentclass[mathleft
]{an}
\usepackage{graphicx}
\usepackage{times}
\usepackage{natbib}
\def\dd{{\rm d}}
\begin{document}

\Pagespan{789}{}
\Yearpublication{}%
\Yearsubmission{}%
\Month{}%
\Volume{}%
\Issue{}%

\title{Seismic signatures of stellar cores of solar-like pulsators:
dependence on mass and age}

\author{I. M. Brand\~ao\inst{1,2}\thanks{\email{isa@astro.up.pt}}, M. S. Cunha\inst{1}, O. L. Creevey\inst{3,4}, J. Christensen-Dalsgaard\inst{5}}

\institute{Centro de Astrof\'isica da Universidade do Porto, Rua das Estrelas, 4150-762 Porto, Portugal
\and
Departamento de Matem\'atica Aplicada, Faculdade de Ci\^encias da Universidade do Porto, Portugal
\and
Instituto de Astrof\'isica de Canarias, E-38200 La Laguna, Tenerife, Spain
\and
Departamento de Astrof\'isica, Universidad de La Laguna, E-38205 La Laguna, Tenerife, Spain.
\and
Department of Physics and Astronomy, Aarhus University, Aarhus, Denmark}

\titlerunning{Seismic signatures of convective cores}
\authorrunning{I. M. Brand\~ao et al.}

\received{??}
\accepted{??}
\publonline{?}

\keywords{stars: oscillations -- stars: interiors}

\abstract{Useful information from the inner layers of stellar pulsators may be derived from the study
of their oscillations. In this paper we analyse three diagnostic tools suggested in the literature
built from the oscillation frequencies computed for a set of main sequence models with masses between $1.0\, {\rm M}_{\odot}$ 
and $1.6\, {\rm M}_{\odot}$, to check what information they may hold about stellar cores. For the models with convective cores ($M \geq 1.2\,{\rm M}_{\odot}$) we find a relation between the frequency slopes of the diagnostic tools and the size of the jump in the
sound speed at the edge of the core. We show that this relation is independent of the
mass of the models. In practice, since the size of the jump in the
sound speed is related to the age of the star, using these seismic tools we may, in principle, infer the star's evolutionary state.
We also show that when combining
two of the three diagnostic tools studied, we
are able to distinguish models
with convective cores from models without a convective core 
but with strong sound-speed gradients in the inner layers
}

\maketitle

\section{Introduction}

The importance of studying the physical properties of the core of a star
is connected to the fact that it is the most determinant region for the
star's evolution. Although stars
with masses similar to the Sun ($M \sim 1\,{\rm M}_{\odot}$) have radiative cores,
those slightly more massive ($M > 1\,{\rm M}_{\odot}$) may develop a 
convective core at some stages of their evolution on the main sequence.
Since convection implies chemical mixing, the 
evolution of these stars is severely influenced by the presence,
and the extent, of the convective core.

Stellar oscillations can provide useful information about the
interior of pulsating stars. For example, at the edge of a convective
region, the derivatives of the sound speed suffer rapid variations that 
leave particular signatures in the oscillation frequencies
\citep{gough90,mjm94,rox94,basu94}.
Several tools have been developed in order to infer 
the properties of the interior of a star.
The most common tools are the large and small 
separations.The first corresponds to the average frequency spacing between modes of 
consecutive radial order but with the same spherical
degree, and provides an estimate of the mean stellar density \citep{cox80}. 
The second corresponds to the 
average spacing between modes of
consecutive radial order but with spherical degrees that differ by 2. It is sensitive to the chemical
composition in the deep interior, tracing the stellar evolutionary state \citep{dappen88}.
The expected signature of large convective cores on the small frequency separations has been considered by 
\cite{rox01} and \cite{rox03}. \cite{cunha07}
carried out a theoretical analysis based on the properties of the oscillations of stellar models slightly more massive
than the Sun and derived the expected signature of a small convective core on the oscillation frequencies.

With the recent launch of the Kepler satellite hundreds of stars will be continuously monitored in search of stellar oscillations 
to a very high degree of precision. So, several questions can be raised: Can we detect the signature of a small 
convective core on the oscillation frequencies 
from photometric data? What would be the dependence of this signature on the stellar mass and physical parameters? What is the precision
required on the individual frequencies in order to infer the signature of 
a convective core? In this work we will address these questions.
\begin{figure*}
\begin{minipage}{\linewidth} 
\begin{center}
\includegraphics[width=8cm]{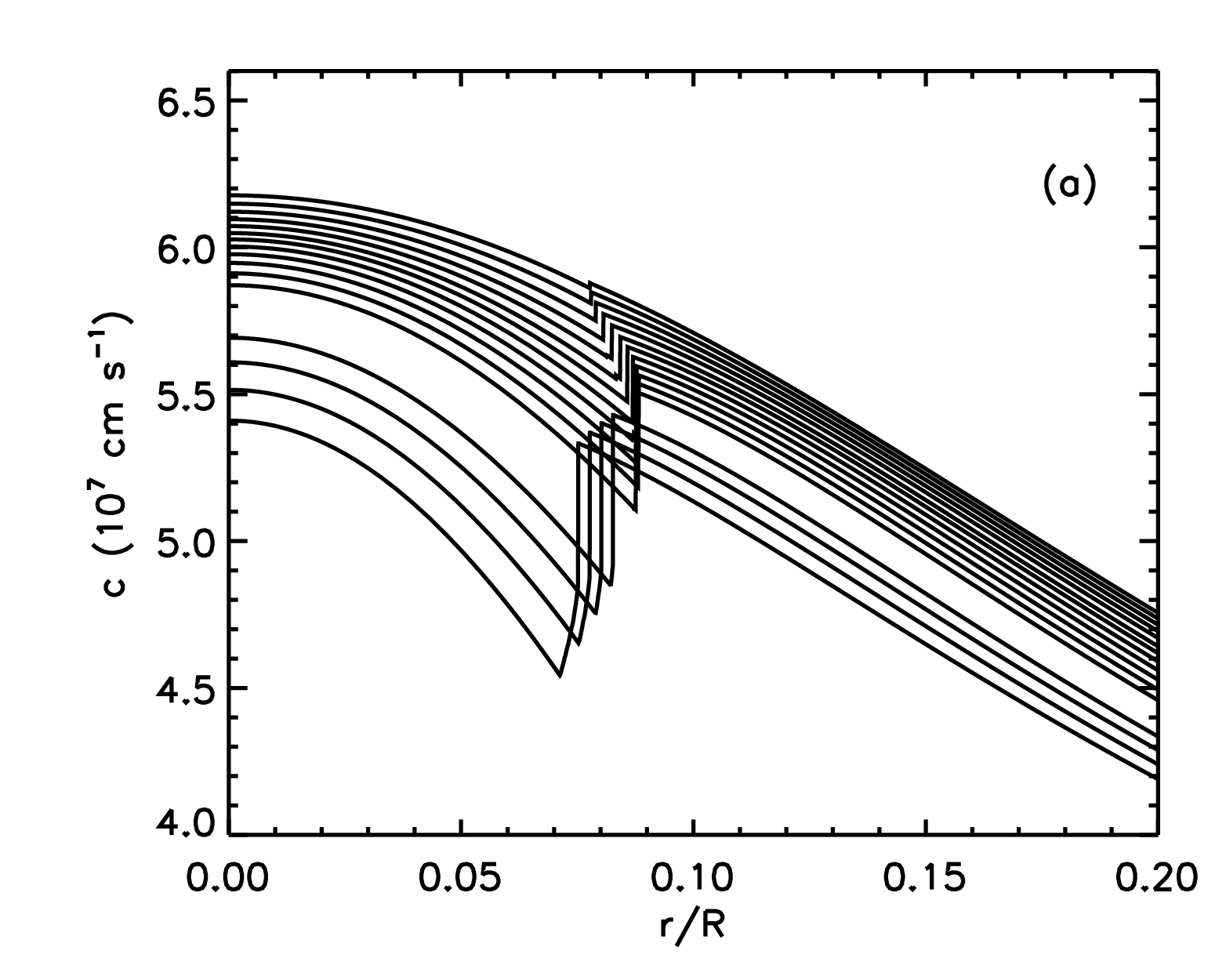} 
\includegraphics[width=8cm]{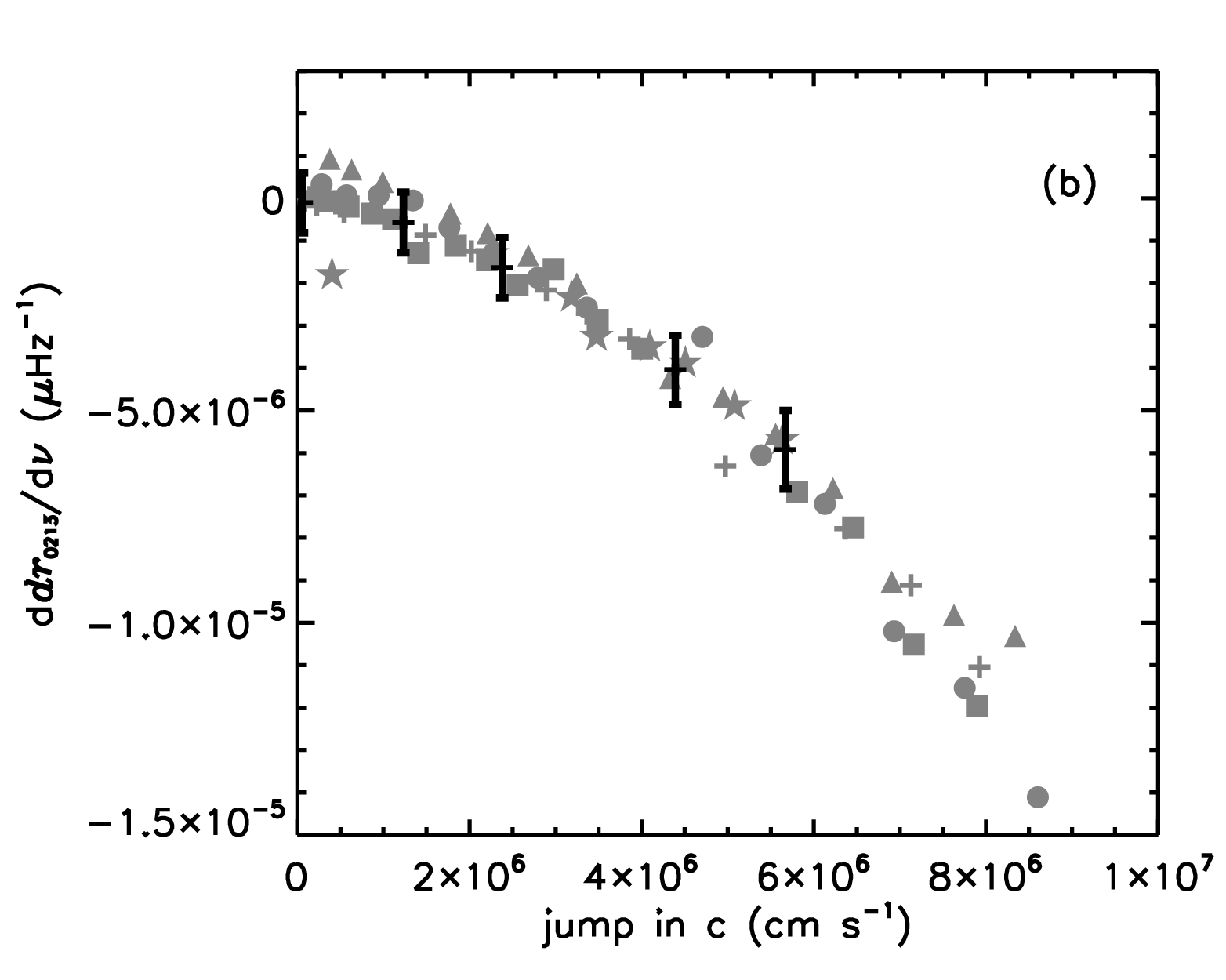}\\
\includegraphics[width=8cm]{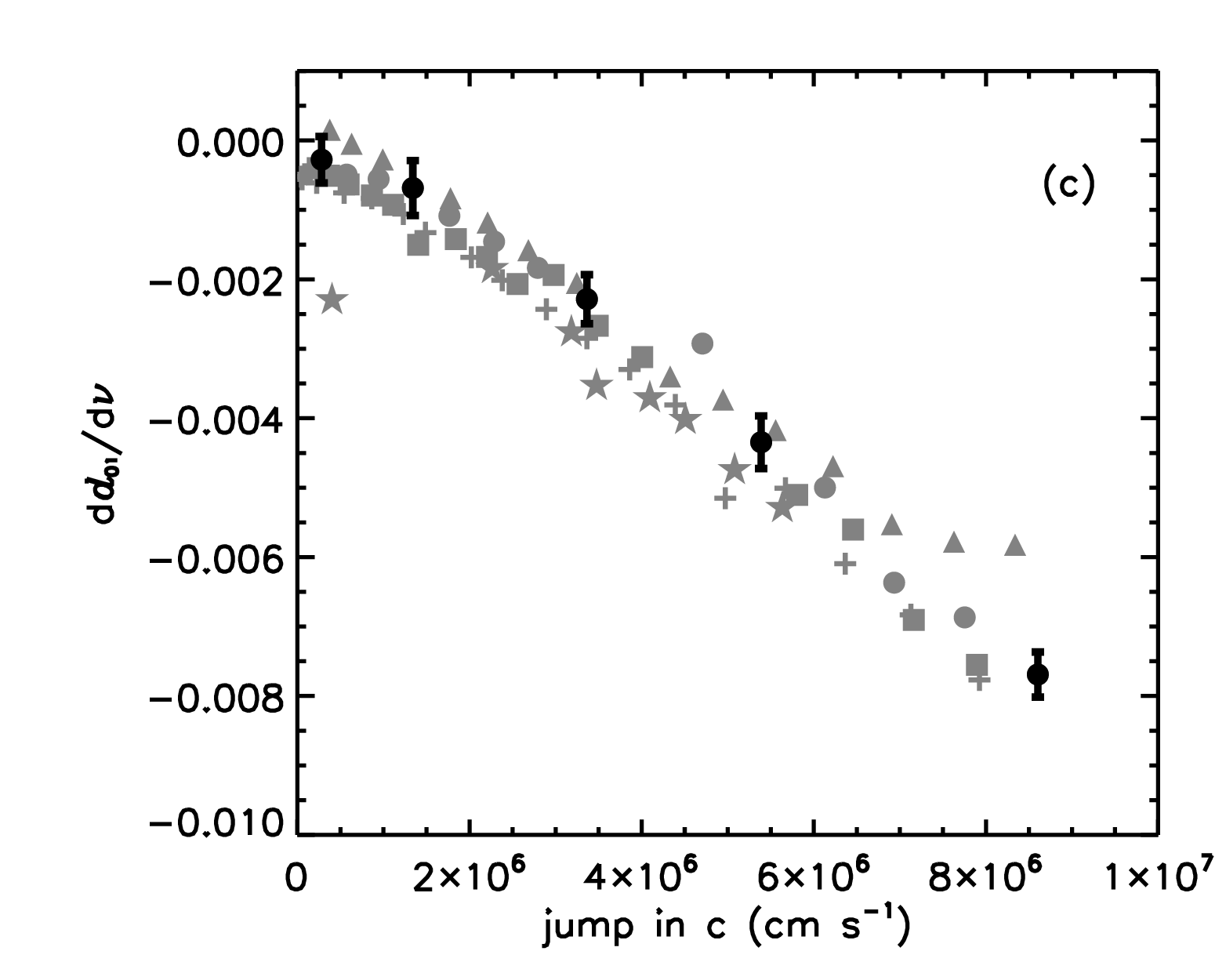}
\includegraphics[width=8cm]{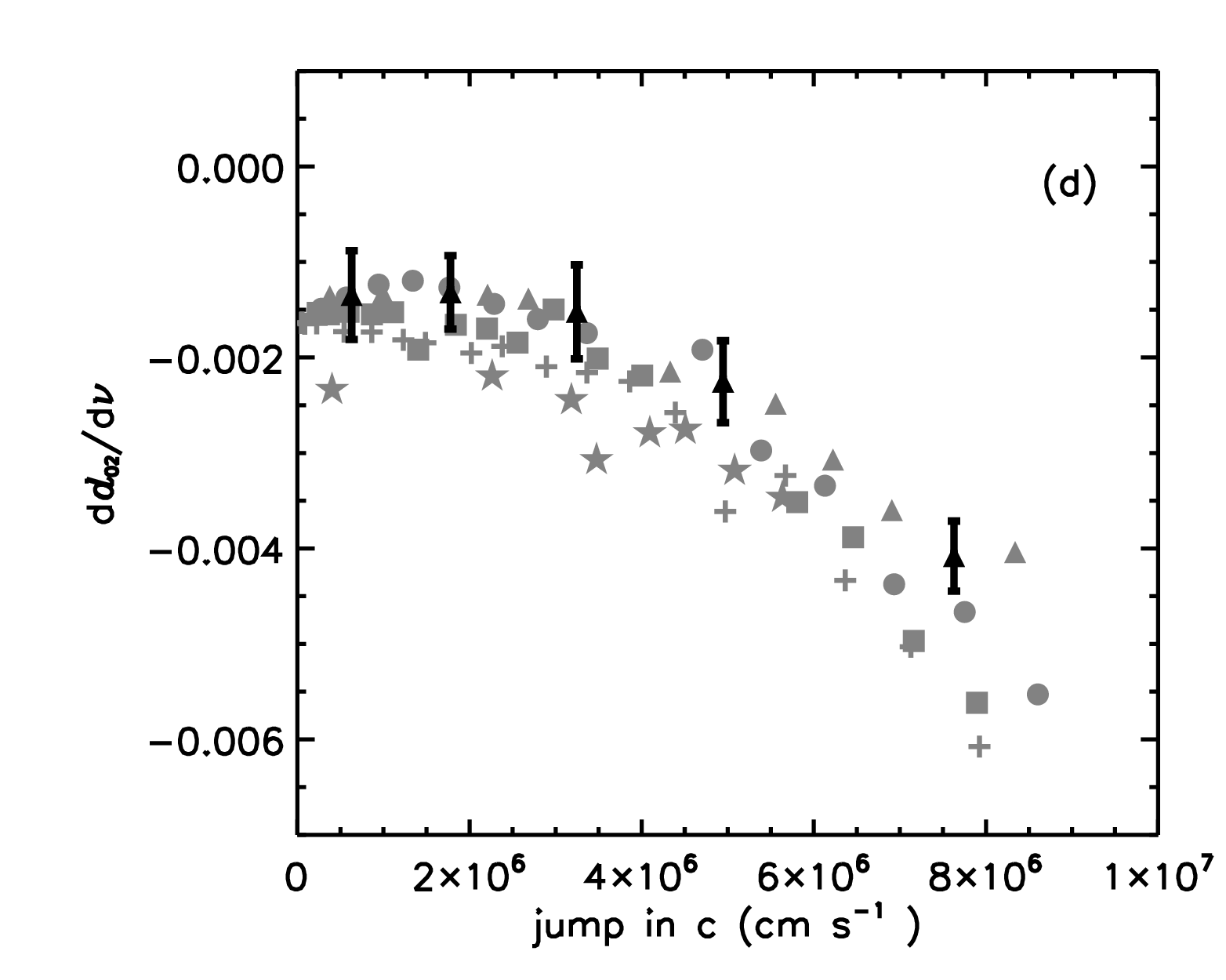}
\end{center}
\caption{Panel (a): sound-speed profile in the
inner layers of a $M=1.4 \, {\rm M}_\odot$ models with 
different ages from 0.2 Gyr (top curve) to 3.6 Gyr (bottom
curve).
The slopes of the $dr_{0213}$ (panel (b)), $d_{01}$ (panel (c))
and $d_{02}$ (panel (d)) diagnostic tools as a function of the
size of the jump in the sound speed at the edge of
the convective cores. Models with masses $M=1.2 \, {\rm M}_\odot$ are represented
by stars, $M=1.3 \, {\rm M}_\odot$ by crosses, $M=1.4 \, {\rm M}_\odot$ by squares, $M=1.5 \, {\rm M}_\odot$ by circles,
and $M=1.6 \, {\rm M}_\odot$ by triangles. The error bars are also plotted for some of the
models.}
\end{minipage}
\end{figure*}

\section{Method}

We started by computing a set of main sequence evolutionary tracks, 
with masses varying from 1.0 to $1.6\, {\rm M}_\odot$ in steps of $0.1 \, {\rm M}_\odot$, 
using the \lq Aarhus STellar Evolution Code\rq, \citep[ASTEC;][]{cd08aastec}.
Core overshoot was included, with an overshoot distance of $\alpha_{\rm ov} = 0.25$, in units of the pressure scale height; the overshoot region was assumed fully mixed and adiabatically stratified.
We assumed an initial helium ($Y$) and
heavy-element ($Z$) abundances of 0.24 and 0.02, respectively.
Diffusion was not considered in this work. The physics used in the code was the same as 
described in \cite{cunha07}.
For some models, at different evolutionary stages,
we computed the oscillation
frequencies using the Aarhus adiabatic oscillation code ADIPLS
\citep{cd08adipls}. These frequencies were then
used to compute the 
following three diagnostic tools:
\begin{equation}
dr_{0213} = \frac{D_{02}}{\Delta \nu_{n-1,1}} - \frac{D_{13}}{\Delta \nu_{n,0}},
\end{equation}
\begin{equation}
d_{01} = \frac{1}{8} (\nu_{n-1,0} -4 \nu_{n-1,1} + 6 \nu_{n,0} - 4 \nu_{n,1}+ \nu_{n+1,0}),
\end{equation}
\begin{equation}
d_{02} = \nu_{n,0} - \nu_{n-1,2},
\end{equation}
where $\nu_{n,l}$ correspond to the model frequencies, for each spherical
degree $l$ and radial order $n$. The first diagnostic tool (Eq.\,1) 
was proposed by \cite{cunha07} and corresponds to a 
difference of ratios
between the scaled small separations,
$D_{l,l+2}\equiv(\nu_{n,l}-\nu_{n-1,l+2})/(4l+6)$, and the
large separations $\Delta\nu_{n,l}\equiv\nu_{n+1,l}-\nu_{n,l}$
for different combinations of mode degrees.
This tool isolates the signature of the core, but
uses modes of degree from 0 to 3. The $l=3$ modes are not easy to
detect from space-based data. For that reason,
we also considered in our study the other two diagnostic tools,
$d_{01}$ (Eq.\,2) \citep[e.g.,][]{rox03} that only involves modes of $l=0$ and 1
and the small separation $d_{02}$ (Eq.\,3)
\citep{gough83} that involves modes of $l=0$ and 2. 
The down side of the latter two diagnostic tools is that 
they do not isolate
the signature of the core,
although they may be strongly affected by it.

In the case of stars with convective cores we expect the frequency slopes 
of the diagnostic tools to be a measure of the jump 
in the sound speed at the edge of the convective core (Cunha et al., these proceedings). 
We plotted each diagnostic tool as a function of frequency
and computed the slope by performing a linear least square fit to 10 frequencies of modes of  
consecutive radial orders, $n$, with $n$ in the range [16,29]. The  
10 radial orders effectively used in the fits were fixed for each  
evolutionary sequence, but varied with stellar mass. In practice,  
they were chosen in a way such as to capture the frequency region in  
which the derivatives of the diagnostic tools (the slopes) were  
approximately constant, and this range is also the expected 
observed range of frequencies.
Moreover, for models with convective cores, we also computed the size of the
jump of the sound speed at the edge of the convective core.
\begin{figure*}
\begin{minipage}{\linewidth} 
\includegraphics[width=8cm]{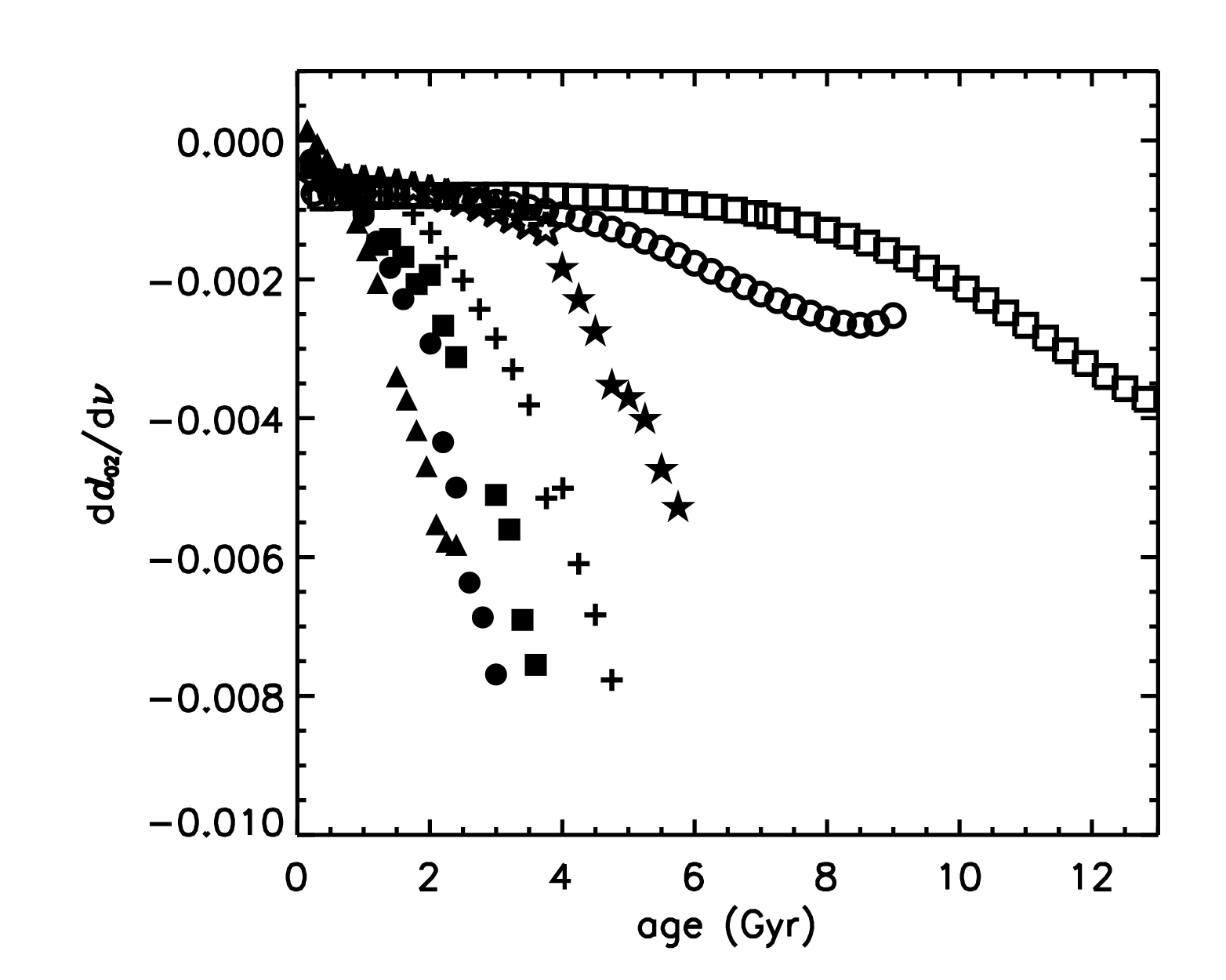}
\includegraphics[width=8cm]{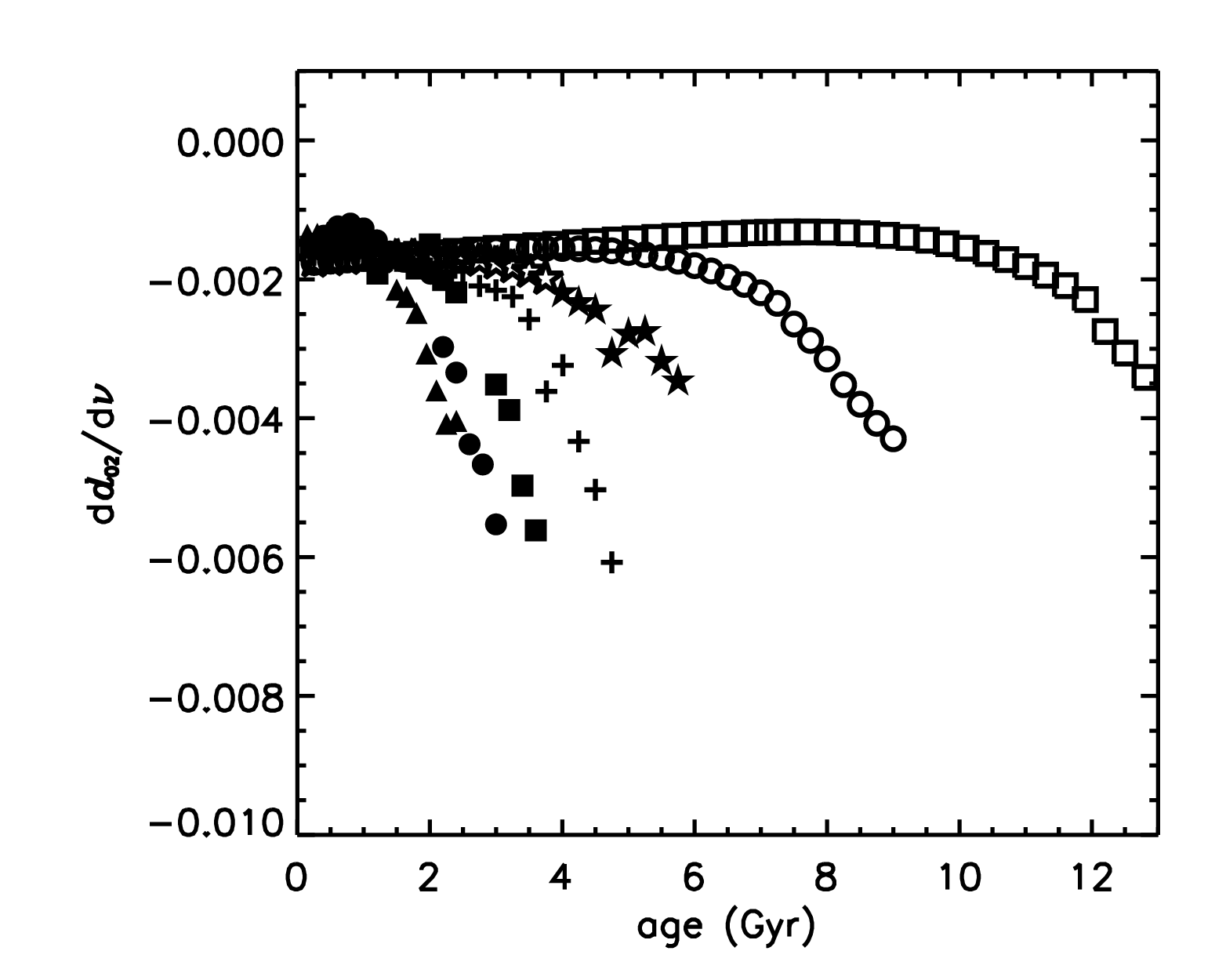}
\caption{The slope of the $d_{01}$ (left plot) and $d_{02}$ (right plot)
diagnostic tools as a function of the age, for all models of our grid.
Open symbols represent models 
without a convective core: squares for $M =1.0 \, {\rm M}_\odot$, circles for $M =1.1 \, {\rm M}_\odot$,
stars for $M =1.2 \, {\rm M}_\odot$ (note that in the latter case, 
models older than 4 Gyr have a convective core).
The other symbols are the same as in Fig.\,1. }
\end{minipage}
\end{figure*}

\section{Results}

We verified that our sequence of $1.0\, {\rm M}_\odot$ models does not show strong sound-speed gradients
in the innermost layers, although the sound-speed gradients are being built 
up as the star evolves. These models do not have a convective core.
Models with $1.1\, {\rm M}_\odot$ also do not have a convective core, but the most evolved
show strong sound-speed gradients that are due
to density variations. Models with $M \geq 1.2\, {\rm M}_\odot$ have convective cores
and show a discontinuity of the sound speed at the core edge (Fig.\,1a). 

Figs.\,1b-d show, for models with a convective core,
slopes of the 
three diagnostic tools considered as a function
of the size of the jump in the sound speed at the edge of 
the core. These three
plots show a clear dependence of the slopes on
the size of the jump, as expected from the work of
Cunha et al. (these proceeding), this dependence 
being independent of the mass, at least for the 
physics that we considered in our set of models. In the second plot, for the
quantity $d_{01}$, we find what seems to be a nearly linear relation.

From these results we can say that, in principle, it is
possible to get a measure of
the size of the jump in the sound speed at the edge of 
a small convective core
from the  
observed oscillation frequencies. However, we are only
able to do that if the uncertainties on the 
observations are such as to allow us
to distinguish between slopes corresponding to 
different jumps.
Considering a relative error
of 10$^{-4}$ on the individual frequencies,
we simulated several sets of frequencies within this error.
We note that observing on long timescales leads to very high
precision frequency spectra.  So even if the mode lifetimes are short
(larger Lorentzian linewidths), the frequencies can
still be determined with accuracy, if the star is "well
behaved" (e.g. not a fast rotator, or a very active star).
We computed the three diagnostic tools for each set of frequencies and measured their slopes.
We considered the 1-$\sigma$ dispersion on the slopes
as an estimate of their uncertainty. The error
bars are shown for some models in the three plots of Fig.\,1
and confirm that one may expect to be able
to set constraints on the jump in sound speed,
particularly when using the $dr_{0213}$ or $d_{01}$
diagnostic tool.
Fig.\,2 shows the slopes of the $d_{01}$ and $d_{02}$ diagnostic tools
as a function of age. As mentioned above, 
although our $1.1\, {\rm M}_\odot$ models do not have
a convective core, they show strong sound-speed gradients in
the core when significantly 
evolved. These gradients also have an impact
on the oscillation frequencies. Thus, we
then ask the question: given a value for the slope, how 
can we distinguish between models with strong sound-speed 
gradients and without a convective core, from those with a 
convective core that show a discontinuity in the sound speed?
To address this question, and since the two
diagnostic tools give different information from the interior
of the star, we plotted the slopes of the $d_{01}$ diagnostic
tool as a function of the slopes of the $d_{02}$ diagnostic
tool (Fig.\,3). In this diagram, in the 
region of small slopes it is hardly possible to distinguish 
between slopes, given the uncertainties. However,
for slopes of magnitude larger than a given treshhold, say 
$\dd d_{02} / \dd \nu <-0.002$, we see that
models with strong sound-speed gradients but no
discontinuity (i.e, no convective core --- open symbols)
are positioned
differently from those with convective cores (filled symbols).
We, thus, see this slope-slope diagram as a 
potentially very interesting tool
to distinguish between stars with strong variations of 
sound speed in the core, yet of different origin.

\section{Conclusions}

We used three seismic diagnostic tools 
to study the cores of models of different
masses and different evolutionary states.
We verified that there is a relation between
the frequency slopes of all of the diagnostic 
tools and the size
of the jump in the sound speed at
the edge of the convective core. In the case 
of the diagnostic tool, $d_{01}$, we find
that the relation is linear. Moreover,
in the case of the $dr_{0213}$ and $d_{01}$ diagnostic tools,
the relation seems to be independent of stellar
mass.
Since strong sound-speed gradients may exist in evolved
models of stars without a convective core, we have checked
if it is possible to distinguish between
those cases and the case of models with
convective cores. 
We found that this may in principle be possible, in some cases, in a
$\dd d_{01} / \dd \nu - \dd d_{02} / \dd \nu$ diagram. Further investigation is underway
to check this possibility.
\begin{figure}
\includegraphics[width=8cm]{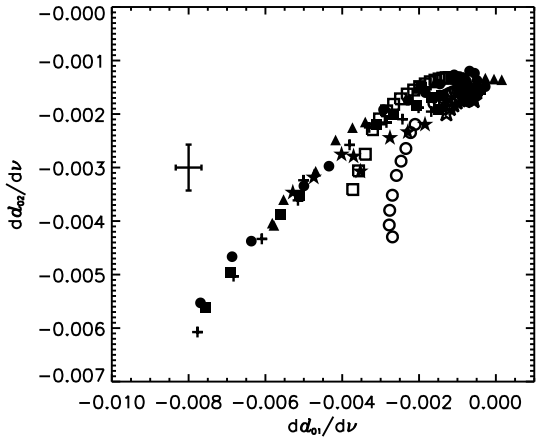}
\caption{The slopes of the $d_{01}$ diagnostic 
tool as a function of the slopes of the $d_{02}$ diagnostic. 
Symbols are the same as for Fig.\,2.}
\end{figure}

\acknowledgements
This work was supported by the European Helio- and Asteroseismology
Network (HELAS), a major international collaboration funded by the
European Commission's Sixth Framework Programme.
Part of this work was carried out during IMB's visit to
the Instituto de Astrof\'isica de Canarias (IAC).
This work was also supported by the project PTDC/CTE-AST/098754/2008, the
grant SFRH / BD / 41213 / 2007 funded by FCT/MCTES, Portugal. 
MC is supported by a Ci\^encia
2007 contract, funded by FCT/MCTES(Portugal) and POPH/FSE (EC).

\newpage


\bibliographystyle{aa}
\bibliography{Isa_lanzarote_final} 


\end{document}